\documentstyle[12pt]{article}

\def\pd{\partial}
\def\a{\alpha}
\def\b{\beta}
\def\dl{\delta}
\def\s{\sigma}

\def\eps{\epsilon}

\def\lam{\lambda}
\def\bg{{\bar g}}
\def\hg{{\hat g}}
\def\hog{{\hat g}_{(\omega)}}
\def\pg{g^{\pp}}
\def\bnabla{{\bar \nabla}}
\def\hnabla{{\hat \nabla}}
\def\tD{{\cal D}}
\def\bR{{\bar R}}
\def\hR{{\hat R}}
\def\bBox{\stackrel{-}{\Box}}
\def\hBox{{\hat \Box}} 
\def\gm{\gamma}
\def\Gm{\Gamma}
\def\om{\omega}
\def\ra{{\rm a}}
\def\sq{\sqrt}
\def\e{\hbox{\large \it e}}
\def\half{\frac{1}{2}}
\def\fr{\frac}
\def\pp{\prime}
\def\arr{\rightarrow}
\def\bb{\begin{equation}}
\def\ee{\end{equation}}
\def\bba{\begin{eqnarray}}
\def\eea{\end{eqnarray}}

\begin{document}

\begin{titlepage}

\begin{tabbing}
   qqqqqqqqqqqqqqqqqqqqqqqqqqqqqqqqqqqqqqqqqqqqqq 
   \= qqqqqqqqqqqqq  \kill 
         \>  {\sc KEK-TH-595 }    \\
         \>       hep-th/9810095 \\
         \>  {\sc October 1998} 
\end{tabbing}
\vspace{5mm}

\begin{center}
{\Large {\bf Background-metric Independent Formulation  
\break of 4D Quantum Gravity}}
\end{center}

\vspace{1.5cm}

\centering{\sc Ken-ji Hamada\footnote{E-mail address : 
hamada@theory.kek.jp} 
and Fumihiko Sugino\footnote{E-mail address : 
sugino@theory.kek.jp}}

\vspace{5mm}

\begin{center}
{\it Institute of Particle and Nuclear Studies, \break 
High Energy Accelerator Research Organization (KEK),} \\
{\it Tsukuba, Ibaraki 305-0801, Japan}
\end{center} 

\vspace{3mm}
\begin{center}
 (Revised Version)
\end{center}
\vspace{3mm}

\begin{abstract} 
Using the background-metric independence for the traceless mode   
as well as the conformal mode, 4D quantum gravity is described 
as a quantum field theory defined on a non-dynamical 
background-metric. The measure then induces an action with 4 
derivatives.  So we think that 4-th order gravity is 
essential and the Einstein-Hilbert term should be treated 
like a mass term. We introduce the dimensionless self-coupling 
constant $t$ for the traceless mode. In this paper we study 
a model where the measure can be evaluated in the limit 
$t \rightarrow 0$ exactly, using the background-metric 
independence for the conformal mode. The $t$-dependence of 
the measure is determined perturbatively using 
the background-metric independence for the traceless mode.  


\end{abstract}
\end{titlepage}

\section{Introduction}
\indent

  Four dimensional quantum gravity~\cite{d}--\cite{bv} 
is one of the most interesting issues left in the 
developments of quantum field theory.   
The big problem in 4D quantum gravity is that the naive 
perturbation theory breaks down. On the other hand  
it is believed that quantum gravity in two dimensions is a 
well-defined quantum field theory~\cite{kpz}--\cite{kn}. 
Certain formulations of 2D quantum gravity have been 
solved exactly~\cite{kpz,dk}. 
This success in two dimensions have inspired many ideas on 
quantum gravity. Based on such ideas conformal mode dynamics 
in 4 dimensions have been studied by Antoniadis, Mazur 
and Mottola~\cite{am,amm1,amm2}. 
In this paper, we develope these ideas further, and re-investigate 
four dimensional quantum gravity including the traceless mode. 

  One of the most important idea to define quantum gravity 
in the generally coordinate invariant way is the 
background-metric independence. 
The original expression of quantum gravity defined by the 
functional integration over the dynamical metric is trivially 
invariant under any change of non-dynamical background-metric. 
But, when the functional measures are re-expressed by ones  
defined on the background-metric, the background-metric 
independence gives strong constraints on the theory. 

  The background-metric independence includes conformal 
invariance, which is just the key ingredient to solve 
2D quantum gravity exactly~\cite{dk}. 
As stressed in ref.~\cite{h} the conformal invariance is purely 
quantum symmetry realized just when gravity is quantized, 
which does not always require the classical theory to be 
conformally invariant. 
Furthermore this idea is independent of dimensions. 
Naively, it is difficult to imagine that 
conformally invariant theory is not well-defined. 
Therefore we think that even in 4 dimensions quantum gravity 
is well-defined if we formulate it in the background-metric 
independent way. In two dimensions it is enough to consider 
the conformal invariance~\cite{dk,h}, while in four dimensions 
it is necessary to consider the background-metric 
independence for the tracelesss mode as well as the conformal mode.   

   In four dimensions the measure induces an action with 4 
derivatives. So we think that the 4-th order action is rather 
natural in 4 dimensions~\cite{dp,wp} and the Einstein-Hilbert 
action should be treated like a mass term~\cite{s}--\cite{bc}, 
where the square of mass is 
the inverse of the gravitational coupling constant. 
The classical limit is then given in the large mass limit.  
 
  The aim of this paper is to give a proper definition of 
4D quantum gravity. In the next section 
we give general arguments about background-metric 
independence before going to concrete calculations.  
In Sect.3 we discuss the induced action for the conformal mode 
in general cases. We here pay attention to the special property 
of D-th order operators in D dimensions~\cite{h}. 
The argument of $D=4$ is essentially used when 
we evaluate the measure for gravitational fields. 
After giving some remarks on the measures of matter fields 
in Sect.4, we evaluate the measure of gravitational field 
in Sect.5.  We then introduce the dimensionless self-coupling 
constant $t$ for the traceless mode and consider the perturbation 
theory on $t$~\cite{kn}. The conformal mode is treated in a   
non-perturbative way. We discuss a model where the measure 
can be evaluated exactly in the $t \arr 0$ limit.  
The model in the limit essentially corresponds 
to the one studied by Antoniadis, Mazur and Mottola~\cite{amm1}   
though their treatment of the $R^2$-term is differnt from ours. 
To evaluate the $t$-dependence we give an ansatz based on the 
background-metric independence for the traceless mode. It is solved 
in self-consistent manner.   
In Sect.6 we give some comments on scaling operators in 4 dimensions.

\section{General Arguments}
\setcounter{equation}{0}
\indent

Quantum gravity is defined by the functional integral over the 
metric field as follows:
\bb 
    Z = \int \fr{[g^{-1}dg]_g [df]_g}{\hbox{vol(diff.)}} 
              \exp \Bigl[-I_{CL}(f,g)\Bigr] ~, 
\ee
where $g$ is the metric field restricted to  
$g_{\mu\nu}=g_{\nu\mu}$ and $\hbox{vol(diff.)}$ is the gauge 
volume for diffeomorphism.  
$f$ is a matter field discussed in Sect.4. 
In this paper we consider scalar and gauge fields.  
The functional measure for integration over the metric is defined by
\bb
    <\dl g, \dl g>_g = \int d^D x \hbox{$\sq g$}g^{\a\b}g^{\gm\dl} 
            (\dl g_{\a\gm} \dl g_{\b\dl}+u\dl g_{\a\b}\dl g_{\gm\dl})~, 
                       \label{2mg}
\ee
where $u>-1/D$ by positive definitness of the norm. 
This definition is rather symbolic because the measure  
depends on the dynamical variables $g$ explicitly. 
The aim of this paper is to rewrite the measure as  
one defined on the non-dynamical background-metric as in 
usual field theories.   

   Decompose the metric into the conformal mode $\phi$, 
the traceless mode $h$ and the background-metric $\hg$ 
as follows:
\bb
    g_{\mu\nu} = \e^{2\phi}\bg_{\mu\nu}  
\ee
and
\bb
     \bg_{\mu\nu}=\bigl( \hg \e^h \bigr)_{\mu\nu}
            = \hg_{\mu\lam} \bigl( \dl^{\lam}_{~\nu} 
                      + h^{\lam}_{~\nu} 
                + \half (h^2 )^{\lam}_{~\nu} + \cdots \bigr) ~. 
\ee
where $tr(h) = h^{\mu}_{~\mu}=0$.  
An arbitrary variation of the metric is given by 
\bb
    \dl g_{\mu\nu} = 2 \dl\phi g_{\mu\nu} 
       + g_{\mu\lam}\bigl(  \e^{-h}\dl \e^h \bigr)^{\lam}_{~\nu} ~. 
\ee
Since $tr ( \e^{-h}\dl \e^h )
 = \int^1_0 ds ~tr ( \e^{-sh}\dl h\e^{sh} )=0$, 
the variation of the conformal mode and that of the traceless mode 
are orthogonal in the functional space defined by 
the norm (\ref{2mg}). 
Therefore the measure of metric can be decomposed as 
\bb
     \fr{[g^{-1}dg]_g}{\hbox{vol(diff.)}}   
       = \fr{[d\phi]_g [\e^{-h}d \e^h]_g}{\hbox{vol(diff.)}}~, 
\ee
where the norms for the conformal mode and the traceless mode 
are defined respectively by 
\bba
    && <\dl \phi, \dl \phi>_g = \int d^D x \sq{g}  
                        (\dl \phi )^2 ~,  
                            \label{2mp}
                  \\ 
   && <\dl h , \dl h >_g = \int d^D x \sq{g} 
                     ~ tr ( \e^{-h}\dl \e^h )^2 ~. 
                            \label{2mh}
\eea

 Let us rewrite the functional measures defined on the dynamical 
metric $g$ into those defined on the non-dynamical 
background-metric $\hg$ as in usual quantum field theories. 
First consider conformal mode 
dependence of the measures. The partition function will be 
equivalently expressed as    
\bb
    Z = \int \fr{[d\phi]_{\hg}[\e^{-h}d\e^h]_{\hg}[df]_{\bg}}{
                  \hbox{vol(diff.)}} 
              \exp \Bigl[ -S(\phi,\bg)-I_{CL}(f,g) \Bigr] ~,
\ee
where $S$ is the action for the conformal mode induced from the 
measures. It is worth making some remarks on this expression.  
The first is that we here do not give any change for the 
classical action. Namely, the induced action 
is purely the contribution from the measures. 
The second is that the measures of metric fields are 
defined on the background metric $\hg$ because of   
$\det \bg = \det \hg$, 
while for matter fields they in general depend on the 
traceless mode explicitly so that they are defined on the 
metric $\bg$.  

   Originally the partition function is defined by 
the metric $g=\e^{2\phi}\bg$ so that the theory should be invariant 
under the simultaneous changes~\cite{dk}:
\bb
      \bg \arr \e^{2\omega}\bg ~, \qquad 
      \phi \arr \phi-\omega~. 
                 \label{2sp}
\ee
In order that the theory is invariant under these changes, 
the action  $S$ should in general satisfy the following 
transformation law:
\bb
   S(\phi-\omega,\e^{2\om}\bg)= S(\phi,\bg)-R(\omega,\phi,\bg) ~. 
                 \label{2wz}
\ee
The measure is then transformed as 
\bba
   &&  [d\phi]_{e^{2\omega}\hg}
       [\e^{-h}d\e^h]_{e^{2\omega}\hg}
       [df]_{ e^{2\omega}\bg}  
             \nonumber \\
   &&   = [d\phi]_{\hg}[\e^{-h}d\e^h]_{\hg}[df]_{\bg}
         \exp \Bigl[ -R(\omega,\phi,\bg) \Bigr] ~.
\eea
Here note that the measure $[d\phi]_{\hg}$ is invariant under 
a local shift $\phi \arr \phi-\om$.   
Because of this property the invariance 
under the changes (\ref{2sp}) 
means the invariance under the conformal change of the 
background: $\hg \arr \e^{2\om}\hg$.    

In this paper we consider the case of 
$R(\om,\phi,\bg)=S(\om,\bg)$, 
which is called the Wess-Zumino condition~\cite{wz}. 
We make some comments on this particular case in the context of 
scalar field. Explicit form of such an action is given in the 
next section. 

  Next consider the background-metric independence for the 
traceless mode. The theory should be invariant under the 
simultaneous changes  
\bb
      \hg  \arr \hg\e^b ~, \qquad 
      \e^h \arr \e^{-b}\e^h ~,
               \label{2sh}
\ee
where $tr (b) =0$, which preserves the combination $\bg= \hg \e^h$. 
The measure for the matter field can be rewritten in the form
\bb
      [df]_{\bg}=[df]_{\hg} \e^{-W(e^h ,\hg)} ~, 
\ee  
where the induced action for the traceless mode should satisfy 
the Wess-Zumino condition~\cite{wz}
\bb
   W(\e^{-b}\e^h ,\hg\e^b )=W(\e^h ,\hg)- W(\e^b ,\hg)~.
                  \label{2w}
\ee
The explicit form of $W$ is discussed in Sect.4.   
Note that the measure $[\e^{-h}d\e^h]_{\hg}$ is left-invariant 
under the change $\e^h \arr \e^{-b}\e^h$ so that the theory 
becomes invariant under the change of the background: 
$\hg \arr \hg\e^b$. Thus the theory becomes invariant under 
any change of the background-metric. 
This is reasonable because the background-metric is quite 
artificial so that the theory should be independent of how to  
choose the background-metric.

\section{D-th Order Operators in D Dimensions}
\setcounter{equation}{0}
\indent 

  Before evaluating the measure of gravitational field, 
we discuss the general cases first. 
Consider $N$ scalar fields $\varphi_A \quad (A=1,\cdots, N)$, 
which have an action with $2n$-th derivatives in D dimensions:   
\bb
    I(\varphi,g)=\fr{1}{2(4\pi)^{D/2}}\int d^D x \hbox{$\sq g$} 
                       \varphi_A D^{(n)}_{AB}\varphi_B  ~,
                    \label{3ia}
\ee
where 
$D^{(n)}_{AB}= (-\Box)^n \dl_{AB} + \Pi_{AB}$ 
is a covariant operator. $\Pi_{AB}$ is a lower-derivative 
matrix operator and $\Box=\nabla^{\mu}\nabla_{\mu}$.   
Let us calculate the induced action $S(\phi, \bg)$ defined 
by the relation
\bb
   \int [d\varphi]_g \e^{-I(\varphi ,g)}  
     = \e^{-S(\phi, \bg)}  
      \int [d\varphi]_{\hg} \e^{-I(\varphi ,g)}~, 
         \label{3a}
\ee
where the functional measure of l.h.s. is defined by 
\bb
     <\dl\varphi, \dl\varphi>_g 
             = \int d^D x \hbox{$\sq g$} 
                  \dl\varphi_A \dl\varphi_A ~.
\ee
The measure of r.h.s. is defined by replacing the determinant of 
metric $\sq{g}$ into $\sq{\hg}$, while note that the action $I$ 
of r.h.s. depends on $g$, not on $\bg$. 
Therefore the argument can be applied to a ``non-conformally'' 
invariant theory also.       

  {}From the definition (\ref{3a}), the variation of the induced 
action for the conformal mode is given by
\bba
   \dl_{\phi} S(\phi,\bg) 
        &=& -\dl_{\phi} \log \hbox{$\det^{-1/2}$} D^{(n)}
            +\dl_{\phi} \log \hbox{$\det^{-1/2}$} \tD^{(n)}
                     \\ 
        &=& \half \int^{\infty}_{\eps}ds 
                Tr \Bigl( \dl_{\phi}D^{(n)} \e^{-sD^{(n)}} \Bigr)
           -\half \int^{\infty}_{\eps}ds 
                Tr \Bigl( \dl_{\phi}\tD^{(n)} 
                  \e^{-s\tD^{(n)}} \Bigr) ~,
                     \nonumber 
                 \label{3e}
\eea
where $\tD^{(n)}=\e^{D\phi}D^{(n)}$ is a non-covariant operator 
and $\eps = 1/L^{2n}$. Here $L \arr \infty$ is a cutoff. 
The variation of the $2n$-th order  operator can be written 
in the form $\dl_{\phi}D^{(n)}=-2n\dl\phi D^{(n)} +\dl K$, 
where $\dl K$ depends on the details of lower derivative terms. 
The variation of $\tD^{(n)}$ is given by 
$\dl_{\phi} \tD^{(n)}=(D-2n)\dl\phi \tD^{(n)}+ \e^{D\phi}\dl K$. 
Using these variations we get the following expression:  
\bba
     \dl_{\phi} S(\phi, \bg) 
      & = & -n Tr \Bigl( \dl\phi \e^{-\eps D^{(n)}} \Bigr) 
             + \half Tr \Bigl( \dl K  D^{(n)-1} \Bigr)  
                \nonumber   \\
       &&   +(n-D/2) Tr \Bigl( \dl\phi \e^{-\eps \tD^{(n)}} \Bigr)
      - \half Tr \Bigl( \e^{D\phi}\dl K  \tD^{(n)-1} \Bigr)  
                \nonumber \\
       & = &  
         -n Tr \Bigl( \dl\phi \e^{-\eps D^{(n)}} \Bigr)
         +(n-D/2) Tr \Bigl( \dl\phi \e^{-\eps \tD^{(n)}} \Bigr) ~.
\eea
The last equality is proved by using the relation between the 
Green functions:  
$<x| \tD^{(n)-1} |x^{\pp}>_{\bg} = <x| D^{(n)-1} |x^{\pp}>_g$ 
such that 
\bba
  &&  Tr \Bigl( \e^{D\phi} \dl K \tD^{(n)-1} \Bigr)
      = tr \int d^D x \sq{\hg} \e^{D\phi} \dl K <x|\tD^{(n)-1}|x>_{\bg} 
               \nonumber \\ 
  && \quad 
      = tr \int d^D x \sq{g} \dl K <x|D^{(n)-1}|x>_g 
      = Tr \Bigl( \dl K D^{(n)-1} \Bigr) ~, 
\eea
where $tr$ takes over the indices $A$, $B$. 

   For $D=2n$, the expression is simplified. 
The non-covariant part vanishes so that the variation of 
the induced action is written by using the covariant quantity 
${\cal H}^{(n)}(x,\eps)=<x|\e^{-\eps D^{(n)}}|x>_g$.  
Furthermore, in this case, the induced action $S(\phi, \bg)$ 
satisfies the Wess-Zumino condition. It is proved in the following. 
Let us apply the simultaneous changes (\ref{2sp}) to both sides 
of the definition ({\ref{3a}). The l.h.s. is invariant under 
the changes so that we obtain the following relation:
\bb
    \e^{-S(\phi-\om, e^{2\om}\bg)} 
        \int [d\varphi]_{e^{2\om}\hg} \e^{-I(\varphi,g)}  
     = \e^{-S(\phi,\bg)} 
        \int [d\varphi]_{\hg} \e^{-I(\varphi,g)} ~. 
\ee
Now define the action $R(\om,\phi,\bg)$ by the relation
\bb
    \int [d\varphi]_{e^{2\om}\hg} \e^{-I(\varphi,g)}  
     = \e^{-R(\om,\phi,\bg)} 
        \int [d\varphi]_{\hg} \e^{-I(\varphi,g)} ~. 
\ee       
Then we obtain the general relation (\ref{2wz}). 
Next consider the variation of $R(\om,\phi,\bg)$ w.r.t. 
the conformal mode $\phi$, which is given by 
\bb
  \dl_{\phi} R(\om,\phi,\bg)  
    =  - \dl_{\phi} \log \hbox{$\det^{-1/2}$} {\cal D}^{(n)}_{\om} 
       + \dl_{\phi} \log \hbox{$\det^{-1/2}$} {\cal D}^{(n)} ~, 
\ee
where ${\cal D}^{(n)}_{\om}= \e^{-D\om}{\cal D}^{(n)}$ 
and ${\cal D}^{(n)}$ has been defined before.  
As in the same way discussed above we obtain the following expression: 
\bb
    \dl_{\phi} R(\om,\phi,\bg) 
     = (D/2 - n) \Bigl[ 
           Tr \Bigl(\dl\phi \e^{-\eps {\cal D}^{(n)}_{\om}} \Bigr) 
           -Tr \Bigl(\dl\phi \e^{-\eps {\cal D}^{(n)}} \Bigr)
                   \Bigr] ~, 
\ee
where we use the relation between the Green functions: 
$<x| {\cal D}^{(n)-1}_{\om} |x^{\pp}>_{e^{2\om}\bg}= 
<x| {\cal D}^{(n)-1} |x^{\pp}>_{\bg}$.  
Therefore, in the case of $D=2n$, the action is independent of 
$\phi$ such that $R(\om,\phi,\bg)=R(\om,\bg)$. 
{}From the condition at $\phi=\om$, the action $R(\om,\bg)$ is nothing 
but $S(\om,\bg)$. Thus we proved that the induced action $S(\phi,\bg)$ 
of $D=2n$ defined by the relation (\ref{3a}) satisfies the Wess-Zumino 
condition.     
 
  In two dimensions consider the usual second order operator.
It is well-known that the finite term of the heat kernel expansion 
for ${\cal H}^{(1)}(x,\eps)$ is given by the scalar curvature.  
Thus the integrated action is given by the Liouville action  
even though the classical theory is not conformally 
invariant~\cite{h}. 

  In four dimensions we must consider the 4-th order operator. 
The induced action is then given by integrating the covariant 
quantity ${\cal H}^{(2)}(x,\eps)$ over the conformal mode. 
{}From the general argument by Duff~\cite{duff}, such a quantity, 
or what is called trace anomaly depends  only on 
two constants $a$ and $b$ so that the induced action is given by
\bba
   S(\phi,\bg) &=& - \fr{1}{(4\pi)^2} \int d^4 x  \int^{\phi}_0 
                       \dl\phi \sq{g} ~\ra^{(2)}_2   
                 \nonumber \\
         &=& \fr{1}{(4\pi)^2} \int d^4 x \int^{\phi}_0 \dl\phi \sq{g} 
                     \biggl[ a \biggl( F+\fr{2}{3}\Box R \biggr) 
                             + b G   \biggr] ~, 
\eea 
where $\ra^{(2)}_2$ is the finite term of 
${\cal H}^{(2)}(x,\eps)$ defined in eq. (\ref{bh}). 
The constant of integration is determined by the condition 
$S(\phi=0, \bg)=0$ because both sides of functional integrations 
are equivalent at $\phi=0$. 
$F$ and $G$ are the square of Weyl tensor and Euler density 
respectively:
\bba
    F &=& R_{\mu\nu\lam\s}R^{\mu\nu\lam\s}-2R_{\mu\nu}R^{\mu\nu} 
           +\fr{1}{3}R^2 ~,
                        \label{3f}
               \\ 
   G &=& R_{\mu\nu\lam\s}R^{\mu\nu\lam\s}-4R_{\mu\nu}R^{\mu\nu} 
           +R^2 ~.
                       \label{3g}
\eea

 The quantities $F$,  $G$ and $\Box R$ are separately integrable 
w.r.t. the conformal mode~\cite{r}. It is useful to consider the 
following combination~\cite{r,am,amm1}:
\bb
      G-\fr{2}{3}\Box R 
         = \e^{-4\phi}\biggl(  
              4 {\bar \Delta}_4 \phi  
            +{\bar G} -\fr{2}{3}\bBox \bR \biggr) ~,    
\ee
where $\Delta_4$ is the conformally covariant 4-th 
order operator defined by  
\bb
      \Delta_4 = \Box^2 
                    + 2 R^{\mu\nu}\nabla_{\mu}\nabla_{\nu} 
                    -\fr{2}{3}R \Box 
                    + \fr{1}{3}(\nabla^{\mu}R)\nabla_{\mu}  
               \label{3d}
\ee
which satisfies $\Delta_4 = \e^{-4\phi}{\bar \Delta}_4$. 
The induced action then becomes
\bba
   && S(\phi, \bg) = 
       \fr{1}{(4\pi)^2} \int d^4x \sq{\hg} \biggl[ 
         a {\bar F} \phi +2b \phi {\bar \Delta}_4 \phi 
     +b \Bigl( {\bar G}-\fr{2}{3} \bBox {\bar R} \Bigr) \phi 
           \biggr] 
                  \nonumber \\ 
   && \qquad\qquad
        -\fr{1}{(4\pi)^2}\fr{a+b}{18} \int d^4 x 
         \Bigl( \sq{g} R^2 -\sq{\hg}{\bar R}^2 \Bigr) ~.
            \label{3s}
\eea
This action really satisfies the Wess-Zumino condition, which can be 
generally proved, if the integrand $\ra^{(n)}_2$ is integrable 
as well as covariant, as follows:
\bba
    &&  S(\phi -\om, \e^{2\om}\bg) 
           = -\fr{1}{(4\pi)^2}\int d^4 x
              \int^{\phi-\om}_0 \dl \s \sq{\hg} 
               \e^{4(\s+\om)} 
             \ra^{(n)}_2 |_{g=e^{2(\s+\om)}\bg} 
                   \nonumber \\ 
    && ~ 
          = -\fr{1}{(4\pi)^2}\int d^4 x
             \int^{\phi}_{\om} \dl \s \sq{\hg} 
               \e^{4\s} \ra^{(n)}_2 |_{g=e^{2\s}\bg} 
          = S(\phi,\bg)-S(\om,\bg) ~. 
\eea
The last equality is proved by dividing the integral region 
$[\om,\phi]$ into $[0,\phi]-[0,\om]$. In the above case 
the first term  rather trivially satisfies the Wess-Zumino 
condition. In the second term the $\sq{g}R^2$-term itself 
does not satisfy the Wess-Zumino condition, but the above 
combination $\sq{g} R^2 -\sq{\hg}{\bar R}^2$ satisfies it.    

  The results of this section are very important when we 
discuss the contributions from the measures of gravity 
in Sect.5.

\section{The Measures of Matter Fields}
\setcounter{equation}{0}
\indent

   In this section we briefly discuss matter field contributions 
to the induced action.  Matter field actions are constructed with  
at most second order derivatives of fields. 
As discussed in the previous section, such fields are rather 
special in 4 dimensions. 
We make some comments on the measures of scalar field and 
gauge field.  

\subsection{Scalar fields}
\indent

  Let us consider scalar field coupled to the curvature as follows: 
\bb
     I_S (X,g) = \half \fr{1}{(4\pi)^2} \int d^4 x 
                  \sq{g} \bigl( g^{\mu\nu}\pd_{\mu}X \pd_{\nu}X 
                           +\xi R X^2 \bigr) ~. 
\ee
{}From arguments of the previous section, the variation of the 
induced action becomes a non-covariant form in this case. 
We now do not know whether such an integrand is integrable 
or not. Even if integrable, the integrated action does not 
satisfy the Wess-Zumino condition so that the theory becomes more 
complicated. 
So we only consider the conformally coupled scalar field with 
$\xi=1/6$, which is described as $I_{CS}$. 

  Instead of the relation (\ref{3a}), we use the following one: 
\bb
   \int [d X]_g \e^{-I_{CS} (X ,g)}
     = \e^{-\Gm(\phi, \bg)} 
      \int [d X]_{\hg} \e^{-I_{CS}(X ,\bg)}~. 
\ee
The difference is that the action $I_{CS}$ of r.h.s. is 
defined on the metric $\bg$, 
not on the metric $g$~\footnote{ 
Note that the conformal invariance of scalar field in 4 dimensions 
is described by rescaling the scalar field as well as the metric as 
$I_{CS}(X,g)=I_{CS}({\bar X},\bg)$, where ${\bar X}=\e^{\phi}X$. Thus 
\bb
     \int [dX]_{\hg} \e^{-I_{CS}(X,g)} 
       \neq \int [dX]_{\hg} \e^{-I_{CS}(X,\bg)}
        = \int [d{\bar X}]_{\hg} \e^{-I_{CS}({\bar X},\bg)}~. 
\ee
}
so that the variation of $\Gm$ is simply given in the form  
$\dl_{\phi} \Gm(\phi,\bg)=-Tr(\dl\phi \e^{-\eps D_{CS}} ) 
+ \half Tr (\dl K_{CS} D_{CS}^{-1})$, 
where $D_{CS}= -\Box +\fr{1}{6} R$ and  
$\dl K_{CS}$ is defined as in the way discussed in Sect.3. This is 
nothing but the definition of the conformal anomaly.  
For conformal coupling the trace including $\dl K_{CS}$ 
vanishes. As a result $\Gm(\phi,\bg)$ is given in the form   
$S(\phi,\bg)$ defined in (\ref{3s}).  
The coefficients $a$ and $b$ in this case  
have already calculated everywhere~\cite{bd,duff} 
\bb 
      a_X= -\fr{N_X}{120}~, \qquad b_X= \fr{N_X}{360} ~, 
                    \label{4x}
\ee
where $N_X$ is the number of conformally coupled scalar fields.

\subsection{Gauge fields}
\indent

  In this subsection we consider abelian gauge fields defined by the 
action
\bb
        I_A(A_{\mu},g) = \fr{1}{4(4\pi)^2} \int d^4 x \sq{g} 
                     g^{\mu\lam} g^{\nu\s} 
                     F_{\mu\nu} F_{\lam\s} ~, 
\ee
where $F_{\mu\nu}=\nabla_{\mu}A_{\nu}-\nabla_{\nu}A_{\mu} 
= \pd_{\mu}A_{\nu}-\pd_{\nu}A_{\mu}$. 
Gauge theory is classically conformally invariant in 4 dimensions 
which is described as $I_A(A_{\mu},g)=I_A(A_{\mu},\bg)$, where 
the gauge field is not rescaled.  
The measure of gauge field is defined by the norm
\bb
    <\dl A,\dl A>_g = \int d^4 x \sq{g} 
               g^{\mu\nu}\dl A_{\mu} \dl A_{\nu} ~.  
\ee
Unlike the case of scalar field it depends on both the 
conformal mode and the traceless mode. 

  As for the conformal mode, it is well known that when we rewrite 
the measure on $g$ into the one on $\bg$, we obtain the induced 
action (\ref{3s}) with the coefficients~\cite{bd,duff}  
\bb   
      a_A = -\fr{N_A}{10}~, \qquad b_A = \fr{31N_A}{180} ~, 
                \label{4a}
\ee
where $N_A$ is the number of gauge fields.   

  Now, consider the induced action for the traceless mode defined by 
\bb
   [dA_{\mu}]_{\bg} = [dA_{\mu}]_{\hg}\e^{-W( e^h ,\hg)} ~.
                     \label{4ma}
\ee
Apply the simultaneous changes (\ref{2sh}) 
in both sides of (\ref{4ma}). 
The measure of l.h.s. (and also the induced action for the conformal 
mode and the classical gauge action) is invariant 
under the changes, while the r.h.s. becomes 
\bb
      [dA_{\mu}]_{\hg e^b}\e^{-W( e^{-b} e^h , \hg e^b )} 
       = [dA_{\mu}]_{\hg}\e^{-W( e^b,\hg) 
                       -W( e^{-b}e^h , \hg e^b )} ~, 
                 \label{4ma2}
\ee
where we use the relation (\ref{4ma}) again with $h$ replaced  
with $b$. The r.h.s. of (\ref{4ma2}) should become the 
original form so that 
$W$ should satisfy the Wess-Zumino condition (\ref{2w}), 
which can be rewritten in more familiar form by introducing 
the one form 
$(V_{\mu})^{\a}_{~\b}=\hg^{\a\lam} \pd_{\mu}\hg_{\lam\b}$ 
and notations $H=\e^h$ and $B=\e^b$ as follows:
\bb
   W(B^{-1}H, V^B_{\mu}) = W(H, V_{\mu}) - W(B, V_{\mu}) 
\ee
where 
\bb
       V^B_{\mu}= (\hg B)^{-1}\pd_{\mu}(\hg B) 
                = B^{-1}V_{\mu} B +B^{-1}\pd_{\mu}B ~.
\ee  

   The solution of the Wess-Zumino condition is 
well-known~\cite{wz}, which is given by  
\bb
    W(H,V_{\mu}) = \zeta \int^1_0 ds \int d^4 x ~
                     tr\bigl( h ~G(V^s_{\mu})\bigr) 
\ee
where $G(V_{\mu})$ is the non-abelian anomaly of the one form 
$V_{\mu}$ and 
\bb
         V^s_{\mu}= \e^{-sh} V_{\mu} \e^{sh} 
                     +\e^{-sh}\pd_{\mu}\e^{sh} ~.
\ee
Thus what remains to do would be to determine the overall coefficient 
$\zeta$, which we do not discuss anymore in this paper.

\section{The Measures of Gravitational Fields}
\setcounter{equation}{0}
\indent

   In this section we consider the measures of conformal and 
traceless modes of gravity. Henceforth we introduce the 
dimensionless self-coupling constant $t$ for the 
traceless mode in the way~\cite{kn} 
\bb
            \bg_{\mu\nu} = (\hg\e^{th})_{\mu\nu} ~. 
\ee
The classical action for the conformal mode is given by 
the $R^2$-action and that for the traceless mode is the Weyl 
action divided by the square of the 
coupling $t$, which is  
\bb
     I_G = \fr{1}{(4\pi)^2} \int d^4 x \sq{g} \Bigl( \fr{1}{t^2}  
        F  +  c  R^2  - m^2 R + \Lambda \Bigr) ~,  
                    \label{5ca}
\ee
where $m^2 $ is the inverse of gravitational constant and 
$\Lambda$ is the cosmological constant. 
In the flat background the 4 derivative 
parts of the Lagrangian have the form 
$\half tr (h \pd^4 h) + 36 c \phi \pd^4 \phi + o(t)$. 
The presence of the Einstein-Hilbert term now gives rise to the 
tachyon problem at the classical level 
for $c>0$ and $\Lambda =0$~\cite{s}--\cite{bc}, 
but in the quantum theory the kinetic term of conformal mode is 
induced from the measures and also we consider $\Lambda \neq 0$ 
case so that such a problem will disappear. 
The question of unitarity still  
remains to be clarified~\cite{s}--\cite{bc}. 
We here only stress that the theory is unitary at the low energy and  
we cannot avoid the 4-th order action to ensure the background-metric independence. 

 The coefficient $c$ is in general arbitrary, 
but for technical reasons it is determined to be a special 
value later.

\subsection{The induced action in the $t \arr 0$ limit}
\subsubsection{Traceless mode}
\indent

  As a first approximation we consider the $t \arr 0$ limit. 
The metric $\bg$ then reduces to the background metric $\hg$ 
so that the matter field, the conformal mode and the traceless 
mode are decoupled each other. So we can evaluate the 
contributions from the measures exactly. 
This approximation is nothing but the one adopted in~\cite{amm1} 
though our management of the $R^2$-terms in eqs. (\ref{3s}) 
and (\ref{5ca}) are different from theirs.  
The difference affects $t$-dependent contributions discussed 
in Sect.5.2.  

  We first calculate the induced action from the measure of 
traceless mode. At the $t \arr 0$ limit the measure (\ref{2mh}) 
divided by $t$ reduces to $[dh]_{\pg}$ defined by the norm 
$<\dl h, \dl h>_{\pg}=\int \sq{\pg}tr (\dl h)^2$,    
where $\pg_{\mu\nu}=\e^{2\phi}\hg_{\mu\nu}$. 
The action now becomes 
\bb
     I^{(0)}_G(h,\pg) 
     = \fr{1}{(4\pi)^2} \int d^4 x \sq{\pg} \Bigl( 
   \half h_{\mu\nu} T(\pg)^{\mu\nu}_{~~,\lam\s} h^{\lam\s} 
        + c R^{\pp 2} -m^2 R^{\pp} +\Lambda \Bigr) ~, 
\ee  
where $h_{\mu\nu}=\pg_{\mu\lam}h^{\lam}_{~\nu}$. 
To justify one-loop calculations we discard the linear term 
of $h$ in the expansion of classical action  
by imposing  the constraints 
$ R^{\pp}_{\mu\nu}= \fr{1}{4} \pg_{\mu\nu}R^{\pp} $ and 
$\nabla^{\pp}_{\mu} R^{\pp} =0 $. 
The induced action is calculated using the quantity 
${\cal H}^{(2)}(x,\eps)$ for the operator $T$,  
or one loop divergence of $\det T$. 

  To calculate the coefficients $a$ and $b$ in (\ref{3s}) 
we have to fix the gauge. 
The Lagrangian for the traceless part is 
described in the form 
\bb
  \half \sq{\pg} h_{\mu\nu} 
      T(\pg)^{\mu\nu}_{~~,\lam\s} h^{\lam\s} 
     = \half \sq{\pg} h_{\mu\nu} 
        T^{NS}(\pg)^{\mu\nu}_{~~,\lam\s} h^{\lam\s} 
        + \chi^{\mu}N(\pg)_{\mu\nu}\chi^{\nu} ~,
\ee  
where 
$\chi^{\mu}=\nabla^{\pp\lam}h^{\mu}_{~\lam}$. 
The nonsigular operator 
$T^{NS}$ and $N$ are defined by eqs.(\ref{at}) and (\ref{an}).   
According to the standard procedure for the 4-th order 
operators~\cite{d,ft1,bc} 
we adopt the gauge-fixing conditon $\chi^{\mu}=0$ and 
gauge-fixing term such that the action $I_G + I_{FIX}$ 
is the only nonsingular action of $T^{NS}$. 
Applying the general coordinate transformation 
$\dl h^{\mu\nu} =\nabla^{\pp\mu} \xi^{\nu} 
                 +\nabla^{\pp\nu} \xi^{\mu}
     - \half (\nabla^{\pp}_{\lam} \xi^{\lam}) g^{\pp\mu\nu}  
$
to the gauge-fixing conditon we obtain the ghost Lagrangian 
$\sq{\pg}\psi^{*\mu}M_{GH}(\pg)_{\mu\nu}\psi^{\nu}$ 
with 
\bb
      M_{GH}(\pg)_{\mu\nu} = \Box^{\pp} \pg_{\mu\nu} 
                    + \half \nabla^{\pp}_{\mu}\nabla^{\pp}_{\nu} 
                    + R^{\pp}_{\mu\nu} ~. 
\ee
Then the contribution from the measure of traceless mode 
can be derived by calculating the quantity 
\bba
    \dl_{\phi} S(\phi, \hg) 
      &=& -\dl_{\phi}  \log 
           \fr{\det^{1/2} N(\pg) \det M_{GH}(\pg)}
              {\det^{1/2} T^{NS}(\pg)} 
          ~\biggl|_{\hbox{kernel part}} 
                     \\
      &=& -\fr{1}{(4\pi)^2} \int d^4 x~ \dl \phi \sq{\pg} \Bigl( 
             \ra^{(2)}_2 (T^{NS}) - \ra^{(1)}_2(N) 
                -2 \ra^{(1)}_2(M_{GH})  \Bigr) ~, 
                \nonumber 
\eea
where $\ra^{(n)}_2$ is defined in eq. (\ref{bh}). 
Using the formulae (\ref{bd}) and (\ref{bq}), 
we obtain the following quantities:
\bba
     \ra^{(2)}_2(T^{NS})
        &=& \fr{21}{10} R^{\pp}_{\mu\nu\lam\s} R^{\pp\mu\nu\lam\s} 
                   + \fr{29}{40} R^{\pp 2} ~, 
                        \\ 
    \ra^{(1)}_2(N) 
       &=& -\fr{11}{180} R^{\pp}_{\mu\nu\lam\s} R^{\pp\mu\nu\lam\s} 
                   + \fr{161}{120} R^{\pp 2} ~, 
                        \\ 
    \ra^{(1)}_2(M_{GH}) 
       &=& -\fr{11}{180} R^{\pp}_{\mu\nu\lam\s} R^{\pp\mu\nu\lam\s} 
                   + \fr{11}{45} R^{\pp 2} ~. 
\eea
The combinations $F^{\pp}$ and $G^{\pp}$ are now described 
in the forms $ R^{\pp}_{\mu\nu\lam\s} R^{\pp\mu\nu\lam\s}
- \fr{1}{6} R^{\pp 2}$ and 
$R^{\pp}_{\mu\nu\lam\s} R^{\pp\mu\nu\lam\s}$ respectively. 
So we can determine the coefficients $a$ and $b$ of the induced 
action, which are given by~\cite{ft1,ft2} 
\bb
     a_h = -\fr{199}{30} ~, \qquad b_h = \fr{87}{20} ~. 
                   \label{5h}
\ee

\subsubsection{Conformal mode} 
\indent

   As in the two dimensional cases~\cite{dk,h}, 
we assume that the contribution 
from the measure of conformal mode is given in the form (\ref{3s}). 
The coefficients $a_{\phi}$ and $b_{\phi}$ are determined in a  
self-consistent way.  
Consider the conformal change of the background metric 
\bb
         \hg_{\mu\nu} \arr \hg^{(\om)}_{\mu\nu} 
                           =\e^{2\om}\hg_{\mu\nu}  ~.
                   \label{5g}
\ee 
 We then obtain the partition function 
\bb
    Z(\hog)= \int [d\phi]_{\hog}
               [dh]_{\hog}[dX]_{\hog}
                [dA]_{\hog} 
          \exp \Bigl[ -{\cal I}^{(0)}(X,A,h,\phi;\hog) 
                                 \Bigr]  
                   \label{5z}
\ee
and 
\bb  
       {\cal I}^{(0)}(X,A,h,\phi;\hog) 
                 = S(\phi, \hog) + I_{CS}(X,\hog) 
                    +I_A(A,\hog)  
                     + I^{(0)}_G(h, \e^{2\phi}\hog)  ~.    
                         \label{5i}
\ee
The coefficients of the induced action are given by 
$a=a_X +a_A + a_h +a_\phi$ and $b=b_X +b_A + b_h +b_\phi$.  
The $\om$-dependence of the measures 
for $X$, $A_{\mu}$ and $h^{\mu}_{~\nu}$ can be 
obtained  by repeating the previous calculations with 
$\phi$ replaced  with $\om$, 
which are given by the action $S(\om,\hg)$ with the 
coefficients (\ref{4x}), (\ref{4a}) and (\ref{5h}) 
respectivily. 

  The contribution from the conformal 
mode is calculated by using the definition of the partition 
function above. 
The Einstein-Hilbert and the cosmological terms have  
dimensional parameters so that, when the 4-th order term exists, 
these terms do not contribute to the 4-th order induced 
action (\ref{3s}).  
To justify calculations we have to set 
the linear term of $\phi$ vanishing. 
To do this, however, we have to relate 
${\hat F}_{(\om)}$ and ${\hat G}_{(\om)}$ so that 
we cannot determine the 
coefficients $a$ and $b$ because for lack of information. 
Therefore we neglect the $\phi^3$-term of the total action.   
It can be carried out by canceling out the $R^2$-terms from the 
classical action and the induced action by taking the value  
\bb 
               c=\fr{1}{18}(a+b) ~. 
\ee
In this case we only calculate the quantity
\bb
    \dl_{\om}S(\om,\hg)
       = - \dl_{\om} \log \hbox{$\det^{-1/2}$}
                    {\hat \Delta}^{(\om)}_4  
       = -\fr{1}{(4\pi)^2} \int d^4 x ~\dl \om \sq{\hog} 
                \ra^{(2)}_2 ({\hat \Delta}^{(\om)}_4) ~. 
\ee
Using the formula (\ref{bd}) and the definition of 
$\Delta_4$ (\ref{3d}), we obtain 
\bb
    \ra^{(2)}_2 ({\hat \Delta}^{(\om)}_4) 
       = \fr{1}{90} \hR^{(\om)}_{\mu\nu\lam\s}
                       \hR_{(\om)}^{\mu\nu\lam\s} 
         + \fr{1}{90} \hR_{(\om)}^2 ~.  
\ee
{}From this we get the values of coefficients in the induced 
action~\cite{amm1}: 
\bb
      a_{\phi}= \fr{1}{15} ~, \qquad b_{\phi}= -\fr{7}{90} ~. 
                    \label{5p}
\ee

\subsubsection{Background-metric independence 
at the $t \arr 0$ limit}
\indent 

  Combining the results calculated before we can extract the 
$\om$-dependence of the measure in the partition function 
(\ref{5z}). We thus obtain the expression 
\bb
    Z(\hog)= \int [d\phi]_{\hg}[dh]_{\hg}[dX]_{\hg}[dA]_{\hg} 
          \exp \Bigl[ -S(\om,\hg)
           -{\cal I}^{(0)}( X^{\om},A,h,\phi;\hog) \Bigr] ~, 
\ee
where $X^{\om}=\e^{-\om}X$ such that 
$I_{CS}(X^{\om},\hog)=I_{CS}(X,\hg)$. 
The coefficients for the induced action $S(\om,\hg)$ and also 
$S(\phi,\hg)$ in the action ${\cal I}^{(0)}$ are given by 
\bba
    && a = -\fr{N_X}{120} -\fr{N_A}{10}
             -\fr{199}{30}+\fr{1}{15} ~, 
                  \label{5a}
                     \\ 
    && b = \fr{N_X}{360}+\fr{31N_A}{180}
             +\fr{87}{20}-\fr{7}{90} ~.  
                  \label{5b}
\eea
Since the measure $[d\phi]_{\hg}$ is now invariant under a local 
shift, it turns out that, changing the variable as  
$\phi \arr \phi -\om$ and using the Wess-Zumino condition 
$S(\phi-\om,\hog) + S(\om,\hg)=S(\phi,\hg)$,  
the partition function goes back to the original form 
defind on the metric $\hg$. 
Thus we proved $Z(\hog)=Z(\hg)$ in the $t \arr 0$ limit.

\subsection{The induced action for $t \neq  0$}
\indent

  The background-metric independence for the traceless mode  
indicates that the $t$-dependence of the induced action, 
apart from $W(\e^h,\hg)$~\footnote{
This action does not affect the later calculations.
}, 
should appear in the combination of 
the metric $\bg=\hg\e^{th}$ 
because the measure defined on the background metric itself is  
invariant under the simultanious changes for the traceless 
mode (\ref{2sh}).   
Now we assume the $t$-dependence of the partition function   
in the following form: 
\bb
    Z 
    =  \int \fr{[d\phi]_{\hg}[\fr{1}{t}\e^{-th}d\e^{th}]_{\hg} 
         [dX]_{\hg}[dA]_{\bg}}{\hbox{vol(gauge)}}  
          \exp \Bigl[ -{\cal I}(X,A,\phi;\bg) \Bigr] ~,
                  \label{5zz}
\ee
where the total action is 
\bba
   &&  {\cal I}(X,A,\phi;\bg) 
              \nonumber \\ 
   && \quad  = \fr{1}{(4\pi)^2} \int d^4 x \sq{\hg} \biggl[ 
                2b(t) \phi {\bar \Delta}_4 \phi 
                         + a(t) {\bar F}\phi 
           + b(t) \Bigl( {\bar G} 
                         - \fr{2}{3}\bBox \bR \Bigr) \phi  
                \nonumber \\ 
   && \qquad
         + \fr{1}{t^2}  {\bar F} 
         +  \fr{1}{18}\bigl( a(t) + b(t) \bigr) \bR^2  \biggr]     
         + \fr{1}{(4\pi)^2} \int d^4 x \sq{g} (-m^2 R + \Lambda ) 
                 \nonumber \\ 
  && \qquad  
           + I_{CS}(X,\bg) + I_A(A,\bg) ~,  
                   \label{5ii}
\eea
where $a(t)=\sum_n a_n t^{2n}$ and $b(t)=\sum_n b_n t^{2n}$ 
with $a_0=a$ and $b_0=b$ given in (\ref{5a}) and (\ref{5b}) 
respectively. 
The coefficient in front of the classical $R^2$-action is now 
defined by the $t$-dependent value 
$c(t)=\fr{1}{18}(a(t)+b(t))$ so that the $R^2$-terms cancel out. 
Here note that the $\bR^2$-term in (\ref{3s}) remains in the 
action. 

  Let us consider the conformal change of the 
background-metric (\ref{5g}). 
The $\om$-dependences of the measures 
are now calculated as perturbations in $t$. 
The contributions from matter fields have 
already been calculated in Sect.3.  
The gravitational contributions are evaluated 
using the total action defined above. 
Expanding the action up to the $t^2$-order, the quardratic terms 
in fields is given by 
\bba
    &&  {\cal I}_2 (\hog)  
        = \fr{1}{(4\pi)^2} \int d^4 x \sq{\hog} \biggl[ 
             \half h_{\mu\nu} \Bigl\{ 
               T^{NS}(\hog)^{\mu\nu}_{~~,\lam\s} 
               + ct^2 \hR^{(\om)} L(\hog)^{\mu\nu}_{~~,\lam\s} 
                       \Bigr\} h^{\lam\s}
                  \nonumber \\  
   && \qquad\qquad\qquad  
          + 2(b+b_1 t^2) \phi {\hat \Delta}^{(\om)}_4 \phi 
          -4(a+b)t \phi \hR_{(\om)}^{\mu\lam\nu\s}
        \hnabla^{(\om)}_{\lam}\hnabla^{(\om)}_{\s} h_{\mu\nu} 
                  \nonumber \\  
   && \qquad\qquad\qquad 
          + \hnabla_{(\om)}^{\lam}h^{\mu}_{~\lam} 
                \Bigl\{   N(\hog)_{\mu\nu} 
        + ct^2 (-\hnabla^{(\om)}_{\mu}\hnabla^{(\om)}_{\nu} 
              +\hR^{(\om)} \hg^{(\om)}_{\mu\nu})  \Bigr\} 
                  \hnabla_{(\om)}^{\lam}h^{\mu}_{~\lam}  
                  \nonumber \\  
   && \qquad\qquad\qquad  
           -\fr{1}{3}a t \phi \hR^{(\om)} 
                 \hnabla^{(\om)}_{\mu}\hnabla^{(\om)}_{\nu}
                           h^{\mu\nu} 
           -\fr{2}{3}b t (\hBox_{(\om)} \phi) 
       \hnabla^{(\om)}_{\mu}\hnabla^{(\om)}_{\nu}h^{\mu\nu}
           \biggr] ~,
\eea 
where $h_{\mu\nu}=\hg^{(\om)}_{\mu\lam}h^{\lam}_{~\nu}$ and   
the second order operator $L^{\mu\nu}_{~~,\lam\s}$ is defined 
by (\ref{al}).  The conditions $\hR^{(\om)}_{\mu\nu}=\fr{1}{4}
\hg^{(\om)}_{\mu\nu}\hR^{(\om)}$ and $\hnabla^{(\om)}_{\mu} 
\hR^{(\om)}=0$ 
are imposed for the linear term of $h$ to vanish. 
Under the conditions, the quartic operator 
${\hat \Delta}^{(\om)}_4$ (\ref{3d}) reduces to the form 
$ \hBox_{(\om)}^2 -\fr{1}{6}\hR^{(\om)} \hBox_{(\om)}$. 

   The gauge-fixing term is defined such that the highest 
derivative terms become diagonal. 
To do this we rewrite the action ${\cal I}_2$ in the form
\bba
    &&    
       \fr{1}{(4\pi)^2} \int d^4 x \sq{\hog} \biggl[ 
             \half h_{\mu\nu} 
           \Bigl\{ T^{NS}(\hog)^{\mu\nu}_{~~,\lam\s} 
           + ct^2 \hR^{(\om)} L(\hog)^{\mu\nu}_{~~,\lam\s} 
                   \Bigr\} h^{\lam\s}
                  \nonumber \\  
   && \qquad\quad  
          + 2(b+b_1 t^2) \phi {\hat \Delta}^{(\om)}_4 \phi 
        + \fr{b^2 t^2}{6} \bigl( \phi \hBox_{(\om)}^2 \phi 
                -\hR^{(\om)} \phi\hBox_{(\om)}\phi \bigr) 
                  \nonumber \\  
   && \qquad\quad 
            -4(a+b)t \phi \hR_{(\om)}^{\mu\lam\nu\s}
      \hnabla^{(\om)}_{\lam}\hnabla^{(\om)}_{\s} h_{\mu\nu} 
            -\fr{1}{3}(a +2b)t \phi \hR^{(\om)} 
      \hnabla^{(\om)}_{\mu}\hnabla^{(\om)}_{\nu}h^{\mu\nu} 
                  \nonumber \\  
   && \qquad\quad 
         + \Bigl(  \chi_{(\om)}^{\mu} 
                +\fr{bt}{2} \hnabla_{(\om)}^{\mu}\phi \Bigr) 
             {\cal N}(\hog)_{\mu\nu}    
             \Bigl(\chi_{(\om)}^{\nu} 
               +\fr{bt}{2} \hnabla_{(\om)}^{\nu}\phi \Bigr)  
           \biggr] ~,
                      \label{5e}
\eea 
where $\chi_{(\om)}^{\mu}=\hnabla_{(\om)}^{\lam} 
h^{\mu}_{~\lam}$ and 
\bb
    {\cal N}(\hog)_{\mu\nu} = N(\hog)_{\mu\nu} 
      + ct^2 (-\hnabla^{(\om)}_{\mu}\hnabla^{(\om)}_{\nu} 
         +\hR^{(\om)} \hg^{(\om)}_{\mu\nu}) ~.
\ee
Thus we take the gauge-fixing term ${\cal I}_{FIX}$  such that 
the last term of the expression (\ref{5e}) disappears  
in the gauge-fixed action ${\cal I}_2 +{\cal I}_{FIX}$. 
This corresponds to take the gauge-fixing condition 
$\chi_{(\om)}^{\mu}+ \fr{bt}{2} \hnabla_{(\om)}^{\mu}\phi =0$. 
The general coordinate transformation 
$\dl g_{\mu\nu}= g_{\mu\lam}\nabla_{\nu}\xi^{\lam} + 
g_{\nu\lam}\nabla_{\mu}\xi^{\lam}$ is expressed as
\bba
   \dl \phi &=& \fr{1}{4} \hnabla^{(\om)}_{\lam}\xi^{\lam} 
                 +\xi^{\lam} \hnabla^{(\om)}_{\lam}\phi  ~,  
               \\
   t\dl h^{\mu}_{~\nu} &=& \hnabla_{(\om)}^{\mu} \xi_{\nu} 
                       +\hnabla^{(\om)}_{\nu} \xi^{\mu}
           - \half \dl^{\mu}_{~\nu}  
                    \hnabla^{(\om)}_{\lam} \xi^{\lam}   
                 + t \xi^{\lam} \hnabla^{(\om)}_{\lam} 
                                 h^{\mu}_{~\nu}   
                        \\ 
       &&
           + \fr{t}{2} h^{\mu}_{~\lam} 
                \Bigl( \hnabla^{(\om)}_{\nu} \xi^{\lam} 
                  - \hnabla_{(\om)}^{\lam} \xi_{\nu} \Bigr) 
            + \fr{t}{2} h^{\lam}_{~\nu} 
                \Bigl( \hnabla_{(\om)}^{\mu} \xi_{\lam} 
                  - \hnabla^{(\om)}_{\lam} \xi^{\mu} \Bigr) 
             + \cdots ~,
                   \nonumber 
\eea
where $\xi_{\mu}= \hg^{(\om)}_{\mu\lam}\xi^{\lam}$.  
Applying it to the gauge-fixing condition 
we can obtain the ghost action. The kinetic term of the 
ghost Lagrangian is now given in the form 
$\sq{\hog}\psi^{*\mu} {\cal M}_{GH}(\hog)_{\mu\nu} \psi^{\nu}$ 
with 
\bb
     {\cal M}_{GH}(\hog)_{\mu\nu} = M_{GH}(\hog)_{\mu\nu} 
             + \fr{b t^2}{8}\hnabla^{(\om)}_{\mu} 
                            \hnabla^{(\om)}_{\nu} ~. 
\ee

   Changing the normalization as    
$\phi^{\pp}= (4b +4b_1 t^2 +\fr{b^2}{3}t^2)^{1/2}\phi$, 
we then obtain the following expression: 
\bb 
    {\cal I}_2 +{\cal I}_{FIX}  
      = \fr{1}{2(4\pi)^2} \int d^4 x \sq{\hog} 
             (\phi^{\pp}, h_{\mu\nu}) {\cal K} 
                \left( \begin{array}{c} 
                         \phi^{\pp} \\ h^{\lam\s} 
                        \end{array} \right) ~,          
\ee  
where 
\bb
      {\cal K} = \left( \begin{array}{cc} 
              \hBox_{(\om)}^2 &     0           \\
      0  & \hBox_{(\om)}^2 \dl^{\mu}_{(\lam}\dl^{\nu}_{\s)} \\
                 \end{array} \right) 
          +\left( \begin{array}{cc} 
                      A &   0             \\
                      0  & C^{\mu\nu}_{~~, \lam\s}  \\
                 \end{array} \right) 
          +  \left( \begin{array}{cc} 
                     0  &   B_{\lam\s}    \\
                     B^{\mu\nu}  &  0     \\
                 \end{array} \right) 
\ee          
and 
\bba
   && A = -\Bigl( \fr{1}{6} + \fr{5}{72}b t^2 \Bigr) 
                 \hR_{(\om)} \hBox_{(\om)} ~, 
                    \nonumber  \\  
   && C^{\mu\nu}_{~~,\lam\s} 
         = T^{NS\pp}(\hog)^{\mu\nu}_{~~,\lam\s}   
           + c t^2 \hR_{(\om)} L(\hog)^{\mu\nu}_{~~,\lam\s} ~, 
                         \\ 
   && B^{\mu\nu} = -\fr{t}{2\sq{b}} \Bigl\{ 
                4(a+b) \hR^{~~(\mu ~\nu)}_{(\om)~\lam~\s}
                 \hnabla_{(\om)}^{\lam}\hnabla_{(\om)}^{\s} 
               +\fr{1}{3}(a+2b) \hR_{(\om)} 
                  \hnabla_{(\om)}^{(\mu}\hnabla_{(\om)}^{\nu)}
                                 \Bigr\}   ~,                
                    \nonumber 
\eea 
where the prime on $T^{NS}$ stands for removing the 
$\hBox_{(\om)}^2$ term. 

  To derive the $\om$-dependence of the measure for gravity 
we have to evaluate the quantity
\bba
     \dl_{\om} S(\om,\hg) 
      &=& - \dl_{\om} \log 
     \fr{\det^{1/2} {\cal N}(\hog) \det {\cal M}_{GH}(\hog)}
               {\det^{1/2}{\cal K}(\hog)}  
          ~ \biggl|_{\hbox{kernel part}} 
                    \\
      &=& -\fr{1}{(4\pi)^2} \int d^4 x ~\dl \om 
               \sq{\hg_{(\om)}} \bigl(  
              \ra^{(2)}_2 ({\cal K}) 
                - \ra^{(1)}_2 ({\cal N}) 
                - 2\ra^{(1)}_2 ({\cal M}_{GH}) \bigr)  ~, 
                   \nonumber
\eea 
Using the generalized Schwinger-DeWitt technique~\cite{bv,ft1} 
summarized in appendix B, we first calculate the divergent part of 
$\log \det {\cal K} = Tr \log {\cal K}$, which 
is expanded in inverse powers of derivatives as 
\bb   
    - Tr \log {\cal K} 
       =  \gm (A) + \gm (C) + \gm(B)   ~, 
\ee
where  
\bba
      \gm (A)
        &=& - 2 Tr \log \hBox_{(\om)} 
            -  Tr \Bigl( A \fr{1}{\hBox_{(\om)}^2}\Bigr)
         + \half Tr \Bigl( A^2 \fr{1}{\hBox_{(\om)}^4} \Bigr)  
                   \nonumber \\
      \gm (C)
        &=&  -2 Tr \log (\hBox_{(\om)} {\rm I}) 
           - Tr \Bigl( C \fr{1}{\hBox_{(\om)}^2} \Bigr) 
           + \half Tr \Bigl( C^2 \fr{1}{\hBox_{(\om)}^4} \Bigr) 
                  \nonumber \\  
      \gm (B)
        &=&   Tr \Bigl( B^2 \fr{1}{\hBox_{(\om)}^4} \Bigr) ~, 
\eea   
where ${\rm I}= \dl^{\mu}_{(\lam}\dl^{\nu}_{\s)}
= \half (\dl^{\mu}_{~\lam} \dl^{\nu}_{~\s} 
+ \dl^{\mu}_{~\s} \dl^{\nu}_{~\lam} )$ and 
$Tr$ includes the trace over the indices $\mu$, $\nu$.   

  The contribution to the induced action from the diagonal part 
of the conformal mode is calculated as   
$
\ra^{(2)}_2(A)= \ra^{(2)}_2({\hat \Delta}^{(\om)}_4)
$, 
where the $t^4$-term is neglected. It turns out that the 
$t^2$-correction does not appear in this part.  
For the diagonal part of the traceless mode we obtain the 
following quantity:
$
    \ra^{(2)}_2 (C) 
      = \ra^{(2)}_2(T^{NS}(\hog))  - 6 c t^2 \hR_{(\om)}^2 ~.  
$
The off-diagonal part is calculated by using the 
formula (\ref{b4}), 
which gives the $t^2$-order contribution 
\bb
    \ra^{(2)}_2 (B) 
      = t^2 \biggl( 
         \fr{(a+b)^2}{4b} \hR^{(\om)}_{\mu\nu\lam\s}
                  \hR_{(\om)}^{\mu\nu\lam\s} 
         - \fr{2a^2+4ab+b^2}{48b} \hR_{(\om)}^2  \biggr) ~.  
\ee
Thus $\ra^{(2)}_2 ({\cal K})$ is given by summing up the 
results from the operators $A$, $B$ and $C$.  
   
  The ghost parts are calculated as 
$
    \ra^{(1)}_2 ({\cal N}) 
      = \ra^{(1)}_2 (N(\hog)) -\fr{5}{2}c t^2  \hR_{(\om)}^2    
$
and also 
$
    \ra^{(1)}_2 ({\cal M}_{GH}) 
      = \ra^{(1)}_2 (M_{GH}(\hog)) 
          -\fr{1}{72}b t^2 \hR_{(\om)}^2  ~.  
$
Combining the above results we finally obtain 
the $t^2$-dependent part of 
$\ra^{(2)}_2({\cal K})- \ra^{(1)}_2({\cal N}) 
- 2 \ra^{(1)}_2({\cal M}_{GH})$ in the form 
\bb
      t^2 \biggl( \fr{(a+b)^2}{4b} 
                  \hR^{(\om)}_{\mu\nu\lam\s}
                     \hR_{(\om)}^{\mu\nu\lam\s} 
                  -\fr{6a^2+40ab+27b^2}{144b} \hR_{(\om)}^2 
                       \biggr) ~. 
\ee

  Now, we can determine the coefficients $a_1$ and $b_1$ from 
the above results. We finally obtain the 
induced action $S(\om, \hg)$ with the coefficients 
\bba
     &&  a_1 = -\fr{6a^2+40ab+27b^2}{24b} ~, 
                   \label{5a1}
              \\
     &&  b_1 = \fr{7}{6}a + \fr{7}{8}b ~.  
                   \label{5b1}
\eea
Here note that the measures of matter fields do not give the 
contributions directly to the coefficients $a_1$ and $b_1$, 
which contribute indirectly to them through the values $a$ and 
$b$ given by (\ref{5a}) and (\ref{5b}). 

  The background-metric independence for the traceless mode 
indicates that the induced action should be in the form 
$S(\om, \bg)$ if it includes the interaction terms. 
As discussed in the previous subsection this $\om$-dependence 
can be removed by changing the field:  
$\phi \arr \phi -\om$ so that the partition function goes back 
to the original one defined on $\hg$ (\ref{5zz}) provided  
$a_1$ and $b_1$  are given by (\ref{5a1}) and (\ref{5b1}). 
In summary we get the coefficients in the action (\ref{5ii}) as 
\bba
   && a(t) = -\fr{N_X}{120} - \fr{N_A}{10}-\fr{197}{30} 
                  \label{5aa}      \\ 
         && \quad + t^2
              \fr{13 N^2_X +1412 N_X N_A + 36988 N_X
                  -7428 N^2_A + 635656 N_A +16011772}
                 {2880(N_X + 62 N_A +1538)} ~, 
                     \nonumber   \\
   && b(t) = \fr{N_X}{360}+\fr{31 N_A}{180}+\fr{769}{180}
            - t^2 \biggl( \fr{7 N_X}{960}-\fr{49 N_A}{1440}
                          +\fr{1883}{480} \biggr) ~. 
                       \label{5bb}
\eea

\section{Discussions on Scaling Operators}
\setcounter{equation}{0}
\indent

  In this paper we proposed the background-meric independent 
formulation of 4D quantum gravity. A model of 4D quantum gravity 
was described as a quantum field theory defined on the 
background-metric (\ref{5zz}) with the coefficients 
(\ref{5aa},~\ref{5bb}) by solving the ansatz of the 
background-metric independence (\ref{5ii}) in the self-consistent 
manner.  

  The problem of renormalizability still remains to be solved,  
but we think that if the diffeomorphism invariance ensures renormalizability,  
our model will be renormalizable because we can easily show that the background-metric independence really ensures the diffeomorphism 
invariance in quantum level~\cite{hf}. 

  The rest of this section is devoted to discuss scaling operators in 4D 
quantum gravity. The cosmological constant and the Einstein-Hilbert 
terms are the lower-derivative operators with conformal charges.  
As in two dimensions, such operators will receive corrections like
\bb
     \Lambda \int d^4 x \sq{\hg} \e^{\a(t)\phi} 
\ee              
for the cosmological constant and 
\bb
    -m^2 \int d^4 x  \sq{\hg}\e^{\b(t)\phi} \Bigl( \bR + \gm(t) 
        \bnabla^{\mu} \phi \bnabla_{\mu} \phi \Bigr)  
\ee
for the Einstein-Hilbert term. 

  Henceforth we take the flat background-metric for simplicity, 
though in perturbation theory we should choose a 
background-metric such that the approximation is well defined.  
At least up to the $t^2$-order, we can use the argument on 
the scaling operators in refs.~\cite{am,amm2}.  
As for the cosmological constant operator, 
the conformal charge is classically given by 
$\a(t)= \hbox{dim}~[\Lambda] =4$. 
In quantum theory it receives a correction as 
$\a(t)=4 +\gm_{\Lambda}$. The anomalous dimension is now calculated 
using the gauge-fixed action in the form 
$\gm_{\Lambda} = \fr{\a(t)^2}{4b^{\pp}(t)}$, where 
$b^{\pp}(t) = b(t)+\fr{b^2 t^2}{12}$, so that the quadratic 
equation is obtained.  Solving the equation, we get the following 
value:  
\bb
    \a(t)=2b^{\pp}(t) \biggl( 1-\sq{1-\fr{4}{b^{\pp}(t)}} \biggr) ~, 
\ee 
where the solution such that $\a(t) \arr 4$ at the 
classical limit $b^{\pp}(t) \arr \infty$ is chosen.
Similarly for the Einstein-Hilbert term we obtain 
\bb
    \b(t)=2b^{\pp}(t) \biggl( 1-\sq{1-\fr{2}{b^{\pp}(t)}} \biggr) ~. 
\ee

  Physically the cosmological constant should be real so that 
we obtain the condition $b^{\pp}(t) \geq 4$ for 4D quantum gravity to 
exist. Recently the evidence of a smooth phase in 4D simplicial 
quantum gravity coupled to $U(1)$ gauge theories is reported 
in the numerical simulations~\cite{bbkptt}, which suggests that 
$b^{\pp}(t) < 4$ for $N_A =0$, but $b^{\pp}(t) > 4$ for $N_A \geq 1$. 
Naively, comparing with our result we obtain the value for the 
coupling constant: $0.11 < t^2 < 0.20$. This seems to indicate  
that the perturbation of $t$ on the flat background-metric is 
not so bad.   

  Finally we give a comment on the consistency of our 
calculations. Consider a constant shift of conformal field: 
$\phi \arr \phi + \eta$. Then the mass scales are rescaled 
as $\Lambda \arr \Lambda \e^{\a(t)\eta}$ and 
$m^2 \arr m^2 \e^{\b(t)\eta}$, and the Weyl action with the 
coefficient $\eta a(t)$ is induced as well.  
This extra Weyl action, however,  gives at most 
$t^4$-corrections to the coefficients $a(t)$ and $b(t)$ 
of the induced action so that our results are self-consistent 
at least up to the $t^2$-order. 
 
\vspace{2mm}

\begin{flushleft}
{\bf Acknowlegments}
\end{flushleft}

The authors wish to thank N. Tsuda for informing us of 
their results~\cite{bbkptt}. We also acknowlege N. D. Hari Dass 
for careful reading of the manuscript. The research of F.S. is 
supported in part by the Japan Society for the Promotion of 
Science under the Postdoctoral Research Program. 

\vspace{5mm}

\begin{center}
{\Large {\bf Appendix}}
\end{center}

\appendix 
\section{Some Important Formulae}
\setcounter{equation}{0}
\indent 

The conformal mode dependence of curvatures is given by
\bba
  R &=& \e^{-2\phi}(\bR - 6\bBox \phi 
                   -6 \bnabla^{\mu}\phi \bnabla_{\mu} \phi ) ~,  
                       \\           
  R_{\mu\nu} &=& \bR_{\mu\nu} -2\bnabla_{\mu}\bnabla_{\nu} \phi 
                 +2\bnabla_{\mu}\phi \bnabla_{\nu}\phi 
                 -\bg_{\mu\nu}(2\bnabla^{\lam}\phi \bnabla_{\lam}\phi 
                                 +\bBox \phi ) ~, 
\eea
where $g_{\mu\nu}=\e^{2\phi}\bg_{\mu\nu}$. 
The square of the Weyl tensor $F$ defined by eq.(\ref{3f}) is 
conformally covariant:
$
        F= \e^{-4\phi}{\bar F} 
$. 
The Euler density $G$ defined by eq.(\ref{3g}) is a total 
derivative, which is proved by using the Riemann identity~\cite{tv}
\bb
     R^{\mu\a\b\gm}R^{\nu}_{~\a\b\gm} -2R^{\mu\a\nu\b}R_{\a\b} 
      -2 R^{\mu\a}R^{\nu}_{~\a} +R^{\mu\nu}R 
     = \fr{1}{4}g^{\mu\nu}G ~. 
\ee 

 The curvature is expanded w.r.t. the traceless mode as
\bba
  \bR &=& \hR -\hR_{\mu\nu}h^{\mu\nu} 
         +\hnabla_{\mu}\hnabla_{\nu} h^{\mu\nu}
         -\fr{1}{4}\hnabla^{\lam} h^{\mu}_{~\nu} 
                             \hnabla_{\lam} h^{\nu}_{~\mu}
               \nonumber \\
      && 
          +\half \hR^{\s}_{~\mu\lam\nu} h^{\lam}_{~\s} h^{\mu\nu} 
          +\half \hnabla_{\nu} h^{\nu}_{~\mu} 
                          \hnabla_{\lam} h^{\lam\mu} 
          -\hnabla_{\mu}(h^{\mu}_{~\nu} \hnabla^{\lam} 
                           h^{\nu}_{~\lam}) 
          + \cdots ~, 
\eea
where $\bg_{\mu\nu}= (\hg \e^h)_{\mu\nu}$ and 
$h_{\mu\nu}=\hg_{\mu\lam}h^{\lam}_{~\nu}$.

  Under the conditions $\hR_{\mu\nu}=\fr{1}{4}\hg_{\mu\nu}\hR$ 
and $\hnabla_{\mu}\hR =0$, the 4-th order actions are expanded 
in the followings. The functions ${\bar G}$ and ${\bar F}$ 
are given by 
\bba
   {\bar G} &=& {\hat G} - 4 \hR^{\mu\lam\nu\s} 
                 \hnabla_{\lam} \hnabla_{\s} h_{\mu\nu} + \cdots ~,
                        \\
   {\bar F} &=& {\hat F} - 4 \hR^{\mu\lam\nu\s} 
                 \hnabla_{\lam} \hnabla_{\s} h_{\mu\nu} 
                 -\fr{1}{3} \hR 
                   ~\hnabla^{\mu}\hnabla^{\nu} h_{\mu\nu}  
                  + \cdots ~. 
\eea
Here note that the linear terms of $h$ can be written in total 
derivative ones because 
$\hnabla^{\mu}\hR_{\mu\a\b\gm}=\hnabla_{\b}\hR_{\a\gm} 
-\hnabla_{\gm}\hR_{\a\b}=0$.  
The $\bR^2$-action is 
\bba
    \int d^4 x \sq{\hg} \bR^2  
     &=& \int d^4 x \sq{\hg} \Bigl( 
           \hR^2 +\half  h_{\mu\nu} \hR
                      L^{\mu\nu}_{~~,\lam\s}(\hg) ~ h^{\lam\s}  
               \nonumber \\
  && \qquad\qquad\quad  
            -\chi^{\mu}\hnabla_{\mu}\hnabla_{\nu} \chi^{\nu} 
            + \hR \chi^{\mu}\chi_{\mu} \Bigr) + \cdots ~, 
\eea
where $\chi^{\mu}= \hnabla^{\lam} h^{\mu}_{~\lam}$ and 
\bb
        L^{\mu\nu}_{~~,\lam\s}(\hg) 
              = \hBox \dl^{\mu}_{(\lam} \dl^{\nu}_{~\s)} 
                   + 2 \hR^{\mu~\nu}_{~\lam~\s} ~. 
                            \label{al}
\ee  
The Weyl action becomes 
\bba
    && \int d^4 x \sq{g} F 
             \nonumber \\
    &&  = 2 \int d^4 x \sq{\hg} \Bigl(     
            \bR^{\mu\nu} \bR_{\mu\nu} -\fr{1}{3} \bR^2 \Bigr)  
                \nonumber \\  
   &&  
       = \int d^4 x \sq{\hg} \biggl[ -\fr{1}{6} \hR^2 
            +\half h_{\mu\nu} T^{NS}(\hg)^{\mu\nu}_{~~,\lam\s} 
                                   h^{\lam\s} 
            + \chi^{\mu} N(\hg)_{\mu\nu} \chi^{\nu} 
                          \biggr]  + \cdots     
\eea
up to the Euler number. 
The nonsingular operators $T^{NS}$ and $N$ are defined by
\bba
     T^{NS}(\hg)^{\mu\nu}_{~~,\lam\s} 
        &=& \hBox^2 \dl^{\mu}_{(\lam} \dl^{\nu}_{~\s)} 
            -\fr{1}{6} \hR\hBox \dl^{\mu}_{(\lam} \dl^{\nu}_{~\s)} 
            + 4 \hR^{\mu~\nu}_{~\lam~\s}\hBox 
                 \nonumber \\
        &&  -\hR(\hnabla^{\mu}\hnabla_{\lam} 
               + \hnabla_{\lam}\hnabla^{\mu}) \dl^{\nu}_{~\s} 
                   \nonumber \\ 
        &&  + 4 \hR^{\mu\a\nu\b} \hR_{\lam\a\s\b} 
            +\fr{1}{4}\hR^2 \dl^{\mu}_{(\lam} \dl^{\nu}_{~\s)} 
            -\fr{4}{3}\hR \hR^{\mu~\nu}_{~\lam~\s} 
                        \label{at}
\eea      
and 
\bb
       N(\hg)_{\mu\nu}= \hBox \hg_{\mu\nu} 
                      -\fr{1}{3}\hnabla_{\mu}\hnabla_{\nu} 
                      -\fr{11}{12} \hR \hg_{\mu\nu} ~. 
                         \label{an}
\ee 

\section{Generalized Schwinger-DeWitt Technique}
\setcounter{equation}{0}
\indent

  Expand the covariant quantity 
${\cal H}^{(n)}(x,s) = <x|\e^{-s D^{(n)}}|x>_g$ 
in 4 dimensions for $n=1,2$ as
\bb
     {\cal H}^{(n)}(x,s) 
          = \fr{1}{(4\pi)^2}\fr{1}{n s^{\fr{2}{n}}} \Bigl(  
             \ra^{(n)}_0 + \ra^{(n)}_n s  
                +\ra^{(n)}_{2n} s^2 + \cdots  \Bigr) ~.
                      \label{bh}
\ee
The $s$-independent term is given by $\ra^{(n)}_2$ for $n=1,2$.  
On the other hand the divergence is described by 
using the quantity $\ra^{(n)}_2$ as follows:
\bb
     - Tr \log D^{(n)} 
      = \fr{\log L^2}{(4\pi)^2} \int d^4 x 
              \sq{g} ~\ra^{(n)}_2  ~. 
                      \label{bra}
\ee
Thus we must calculate the $\log L^2$ divergences to determine 
the coefficients $\ra^{(n)}_2$ up to the $\Box R$-term.    

  Let us first consider the 4-th order operator in 4 dimensions 
defined by eq.(3.1) with $n=2$. It is here expressed as 
\bb
        D = \Box^2 {\rm I} + \Pi ~, 
\ee
where ${\rm I}$ is the identity operator for the indices $A$, $B$ 
and $\Pi$ is at most second order matrix operator. 
Then one gets the following expression~\cite{bv}: 
\bb
    - Tr \log D  
    = - 2 Tr \log \Box {\rm I}
       - Tr \biggl( \Pi \fr{1}{\Box^2} \biggr) 
         +\half Tr \biggl( \Pi^2 \fr{1}{\Box^4} \biggr) ~. 
\ee
The r.h.s. is calculated by using the universal functional trace 
formulae~\cite{bv}
\bba
  && Tr \log \Box {\rm I} |^{div} 
      = - \fr{\log L^2}{(4\pi)^2} \int d^4 x \sq{g}~ 
           tr \biggl[ \biggl( \fr{1}{180} 
                R_{\mu\nu\lam\s} R^{\mu\nu\lam\s} 
                   \nonumber \\ 
  && \quad\qquad\qquad\qquad\quad  
              -\fr{1}{180} R_{\mu\nu} R^{\mu\nu} 
              +\fr{1}{72}R^2 \biggr){\rm I} 
               +\fr{1}{12} {\cal R}_{\mu\nu}{\cal R}^{\mu\nu} 
                         \biggr] ~ 
\eea  
and 
\bba
        \nabla_{\mu}\nabla_{\nu} \fr{{\rm I}}{\Box^2} 
                  \dl(y,x)|^{div}_{y=x}
          &=& \fr{\log L^2}{(4\pi)^2} \sq{g} 
                   \biggl[ \fr{1}{6} \biggl( 
                   R_{\mu\nu} -\half g_{\mu\nu} R \biggr) {\rm I} 
                   + \half {\cal R}_{\mu\nu} \biggr] ~, 
                        \\   
      \nabla_{\mu}\nabla_{\nu}\nabla_{\lam}\nabla_{\s}
              \fr{{\rm I}}{\Box^4} \dl(y,x)|^{div}_{y=x}
          &=& \fr{\log L^2}{(4\pi)^2} \sq{g} 
               \fr{{\rm I}}{24} \bigl( g_{\mu\nu}g_{\lam\s}
                           +g_{\mu\lam}g_{\nu\s} 
                           +g_{\mu\s}g_{\nu\lam} \bigr) ~,  
                       \label{b4}
\eea 
where $({\cal R}_{\mu\nu} )^A_{~B}$ is defined by 
\bb
        [ \nabla_{\mu}, \nabla_{\nu}]\varphi^A 
          = ({\cal R}_{\mu\nu})^A_{~B} \varphi^B ~. 
\ee

  It is useful to consider the following general form 
to evaluate the diagonal parts of gravity sector in Sect.5:  
\bb
     D = \Box^2 {\rm I} + X^{\a\b}\nabla_{\a} \nabla_{\b} 
               + Y^{\a} \nabla_{\a} + Z ~,  
\ee
where $X^{\a\b}=X^{\b\a}$. 
Using the formulae listed above one gets the following expression 
for the divergent part~\cite{ft1,bv}:    
\bba
     - Tr \log D |^{div} 
      &=& \fr{\log L^2}{(4\pi)^2} \int d^4 x \sq{g}~ 
             tr \biggl[ \fr{1}{90} \Bigl( 
                          R_{\mu\nu\lam\s} R^{\mu\nu\lam\s} 
                          - R_{\mu\nu} R^{\mu\nu} 
                              \Bigr) {\rm I} 
                      \nonumber \\ 
      && \qquad\qquad 
                        +\fr{1}{6} {\cal R}_{\a\b} 
                              {\cal R}^{\a\b} 
                        +\fr{1}{36} R^2 {\rm I} -Z 
                        -\fr{1}{6} R_{\a\b} X^{\a\b} 
                       \nonumber \\ 
     && \qquad\qquad
                        +\fr{1}{12} R X
                        +\fr{1}{48} X^2 
                        +\fr{1}{24} X_{\a\b} X^{\a\b} 
                           \biggr] ~, 
                      \label{bd}
\eea
where $X=X^{\a}_{~\a}$. 

  In the case of $\varphi_A = h_{\mu\nu}$, the quantities 
$X$, $Y$ and $Z$ can be read from the expressions in 
appendix A and   
$
     ({\cal R}_{\a\b})^{\mu\nu}_{~~\lam\s} 
      = R^{\mu}_{~\lam\a\b} \dl^{\nu}_{~\s} 
        + R^{\nu}_{~\s\a\b} \dl^{\mu}_{~\lam} 
$. 
In calculating the trace for the indices $A=(\mu\nu)$, 
one has to take the traceless condition for $h_{\mu\nu}$ into 
account.    
It is carried out by replacing $h_{\mu\nu}$ with $H_{\mu\nu}$ 
defined by the relation 
$h_{\mu\nu}=H_{\mu\nu}-\fr{1}{4}g_{\mu\nu}H^{\lam}_{~\lam}$, or 
replacing the operator $(X^{\a\b})^{\mu\nu}_{~~,\lam\s}$, 
for example, with 
$
({\rm I}_H X^{\a\b} {\rm I}_H )^{\mu\nu}_{~~,\lam\s}
$, 
where 
$
   ({\rm I}_H)^{\mu\nu}_{~~,\lam\s}  
      = \dl^{\mu}_{(\lam} \dl^{\nu}_{\s)} 
          -\fr{1}{4}g^{\mu\nu}g_{\lam\s} 
$ and ${\rm I}^2_H = {\rm I}_H $. 
  
   To evaluate the ghost parts it is useful to consider 
the following second order operator:  
\bb
      Q_{\mu\nu} = \Box g_{\mu\nu} 
                   -\lam \nabla_{\mu} \nabla_{\nu} 
                   + Z_{\mu\nu} ~.  
\ee
The divergent part is given by the formula~\cite{ft1,bv}
\bba
     -Tr \log Q |^{div} 
      &=& \fr{\log L^2}{(4\pi)^2} \int d^4 x \sq{g}~ \fr{1}{3} 
           \biggl[ -\fr{11}{60} \biggl(
                   R_{\mu\nu\lam\s} R^{\mu\nu\lam\s} 
                   -4 R_{\mu\nu} R^{\mu\nu} + R^2 \biggr) 
                \nonumber \\ 
      && \qquad\qquad 
                   +\biggl( 
                       \fr{\gm^2}{8}+\fr{\gm}{4}-\fr{4}{5} 
                         \biggr) R_{\mu\nu} R^{\mu\nu} 
                   +\biggl( 
                       \fr{\gm^2}{16}+\fr{\gm}{4}+\fr{7}{20} 
                         \biggr) R^2
                  \nonumber \\ 
      && \qquad\qquad 
                   + \biggl( 
                       \fr{\gm^2}{4}+\gm
                         \biggr) R_{\mu\nu} Z^{\mu\nu} 
                   + \biggl( 
                       \fr{\gm^2}{8}+\fr{3}{4}\gm +\fr{3}{2} 
                         \biggr) Z_{\mu\nu} Z^{\mu\nu} 
                \nonumber \\  
      && \qquad\qquad
                   + \biggl( 
                       \fr{\gm^2}{8}+\fr{\gm}{4}+\half 
                         \biggr) R Z  
                   + \fr{\gm^2}{16} Z^2 \biggr] ~, 
                         \label{bq}
\eea
where $\gm =\lam/(1-\lam)$ and $Z=Z^{\mu}_{~\mu}$.

\end{document}